\documentclass[12pt]{iopart}
\usepackage{iopams}
\usepackage{epsfig,cite}
\newcommand{\be}{\begin{equation}}
\newcommand{\ee}{\end{equation}}
\newcommand{\bd}{\begin{displaymath}}
\newcommand{\ed}{\end{displaymath}}
\newcommand{\BE}{\begin{eqnarray}}
\newcommand{\EE}{\end{eqnarray}}

\newcommand{\avg}[1]{\left\langle{#1}\right\rangle}

\newcommand{\req}[1]{(\ref{#1})}

\newcommand{\eps}{\epsilon}

\begin{document}

\title{Dynamical instabilities in a simple minority game with discounting}

\author{D Challet\P, A De Martino\dag ~and M Marsili\ddag}

\address{
\P\ Institute for Scientific Interchange (ISI), Viale S. Severo 65, 10133 Torino (Italy) and D\'epartement de Physique, Universit\'e  de Fribourg, P\'erolles, 1700 
Fribourg, Switzerland\\
\dag\ CNR/INFM SMC, Dipartimento di Fisica,
  Universit\`a di Roma ``La Sapienza'', p.le A. Moro 2, 00185 Roma (Italy)\\
\ddag\ The Abdus Salam International Centre for Theoretical Physics, Strada Costiera 14, 34014 Trieste (Italy)\\}

\begin{abstract}
  We explore the effect of discounting and experimentation in a simple model of interacting adaptive agents. Agents belong to either of two types and each has to decide whether to participate a game or not, the game being profitable when there is an excess of players of the other type. We find the emergence of large fluctuations as a result of the onset of a dynamical instability which may arise discontinuously (increasing the discount factor) or continuously (decreasing the experimentation rate). The phase diagram is characterized in detail and noise amplification close to a bifurcation point is identified as the physical mechanism behind the instability.
\end{abstract}

\pacs{02.50.Le, 87.23.Ge, 05.70.Ln, 64.60.Ht}

\ead{\tt challet@isi.it, andrea.demartino@roma1.infn.it, marsili@ictp.it}

\section{Introduction}

Individual optimizing behavior is sometimes sufficient for a
population of agents to coordinate on efficient outcomes. For example,
competitive behavior in markets allows buyers and sellers to
coordinate their demand and supply, and drivers trying to minimize
transit time may avoid crowded routes, thus reducing the chances of
traffic jams. Such consequences of individual behavior have been
thoroughly studied in economics within an equilibrium framework. In
many situations, however, such idealized conditions as those
postulated in economics (e.g. perfect information and rationality),
may be unrealistic \cite{ElFarol}. When individuals are engaged in
contexts involving many other individuals, as in financial markets, it
is normally more realistic to assume that they acquire information
from their environment and from other agents as they learn the
relative efficiency of their strategies/actions. Experimentation --
i.e. the fact of sometimes playing sub-optimal actions -- and
flexibility -- the ability to change strategy in response to a
changing environment -- become important aspects of a learning
dynamics in an evolving system. The former means that agents' behavior
is best represented by a stochastic choice model \cite{Logit}.
Flexibility implies instead that agents should discount past payoffs
in their learning behavior, as they might not be relevant for the
current state of their environment (at the same time, discounting too
much past payoffs, may not allow agents to learn the full complexity
of their environment \cite{MMRZ}). It is thus essential to understand
how the efficiency of learning is affected by individual stochastic
choice and by payoff discounting.

Here we address these issues in an extremely simple setting that
allows us to draw sharp conclusions and to achieve a full
understanding of the results. The model we consider is a stylized
description of situations where two different groups of agents, taken
of equal size for the sake of simplicity, interact, each of them
providing a resource for agents of the other type. Examples of this
generic setting include buyers and sellers -- which is the situation
we shall informally refer to henceforth -- but also (heterosexual)
individuals of opposite sex attending a party: for each of them the
party might be most interesting if the majority of the participants
are of the opposite sex. In addition to this, individuals may have
{\em a priori} incentives to participate in the game. In our model,
individuals take decisions based on an estimate of the payoff, which
they compute from the outcomes of the game in the recent past.

The model falls in the class of Minority Games (MGs) \cite{MG}, which
have been studied extensively in the recent past. Most theoretical
results have been confined so far to the case of infinite memory,
where agents do not discount payoffs. Numerical simulations for more
complex versions of the model than that discussed here show that,
interestingly, strong fluctuations can arise when the discount factor
is introduced \cite{MMRZ,finitemem}. Remarkably, these fluctuations
arise after a much longer time than any characteristic time scale in
the system. Understanding the origin of such non-trivial fluctuations
is indeed one of the motivations of the present work.

Strong fluctuations and dynamical instabilities are persistent and
ubiquitous in a multitude of social systems that can be modeled with
interacting adaptive agents (like economies, financial markets,
traffic, elections, web-communities, etc.; see e.g.
\cite{Bouchaud,Marsili}), hence the issues raised here have a rather
general relevance. Dynamic instabilities in models of interacting
agents have been also addressed in the economics literature (see
Ref. \cite{Hommes} for a review), but the role of stochastic
fluctuations has only been recently recognized \cite{Lux}. 

The aim of the present paper is to present a model which is simple
enough to allows us to understand the origin of strong fluctuations in
a system of boundedly rational interacting agents. In particular we
will focus on the interplay between discounting and the stochastic
nature of the learning process.

We find that, in a region of parameter space, increasing the discount
factor, the system crosses a discontinuous transition and enters a
region where two distinct dynamical steady state solutions exist. The
emergence of strong fluctuations is then an activated process and, as
such, these materialize (and dematerialize) after times which can be
extremely long. This same transition arises continuously, decreasing
the randomness (or experimentation) in the choice behavior, at a
critical threshold. At a still higher threshold a stable solution
reappears, coexisting with the strongly fluctuating one. 

\section{The model}

We consider a system of $N$ buyers and $N$ sellers. We denote by $B$ and $S$ the sets of buyers and sellers, respectively. At time step $t$, agent $i$ can either play the game ($n_i(t)=1$) or abstain ($n_i(t)=0$). We assume that he chooses to play with probability 
\be {\rm
  Prob}\{n_i(t)=1\}=\frac{1+\tanh[\Gamma U^{\pm}_i(t)]}{2} 
\ee 
where
$\Gamma\geq 0$ is the (uniform) learning rate of agents, while the functions
$U_i^{\pm}(t)$, representing the cumulated score of buyers and
sellers, respectively, evolve according to 
\be
\label{iaia}
U_i^{\pm}(t+1)=(1-\lambda)U^{\pm}_i(t)\mp A(t)-\eps 
\ee 
Here $\epsilon$ is a cost (or incentive if $\epsilon<0$) which agents incur for participation. This represents outside opportunities, such as investing in a risk free asset instead of buying. $A(t)$ stands for the participation imbalance (or `excess demand'), defined as 
\be
A(t)=\frac{1}{N}\left[\sum_{i\in B} n_i(t)-\sum_{i\in S}n_i(t)\right].
\ee 
Finally the constant $\lambda$ ($0\leq\lambda\leq 1$) plays the
role of a discount factor, introducing a finite memory on the score. 
In words, agents estimate a score for participating in the game taking an exponential moving average of a payoff. The payoff $\mp A(t)-\epsilon$ depends on how many more agents of the other type had recently participated (i.e. on $A(t)$) and on a constant cost $\epsilon$. Agents base their decision on
whether to play or not by learning how profitable was it to
play in the past. Note that the reinforcement
term in the learning dynamics is independent of the agent, hence we
can drop the index $i$ from \req{iaia}. 

For the analysis which follows,  it is reasonable to separate
the sum and the difference of $U^{\pm}$ by introducing the variables
\BE
\bar u(t)=\Gamma\frac{U^+(t)+U^-(t)}{2}\\
y(t)=\Gamma\frac{U^+(t)-U^-(t)}{2}
\EE
so that $U^\pm=(\bar u\pm y)/\Gamma$. The dynamics of $\bar u$ is
deterministic and it can be easily solved to yield
\be
\label{Ubar}
\bar{u}(t)=\left[\bar{u}(0)+\frac{\Gamma\epsilon}{\lambda}\right](1-\lambda)^t-\frac{\Gamma\epsilon}{\lambda}
\ee
so that $\bar u$ converges over times of the order of $1/\lambda$ to its asymptotic value
$-\Gamma\eps/\lambda$. We shall henceforth neglect such
a transient, and set $\bar u=-\Gamma\eps/\lambda$. The time evolution of $y$ is instead described
by 
\be
\label{dyny}
y(t+1)=y(t)(1-\lambda)-\Gamma A(t).
\ee
In order to understand the structure of fluctuations, we shall now focus on the above process.

\section{Deterministic approximation}

For $N\to\infty$ a deterministic theory holds. Indeed the quantity $A(t)$ satisfies the law of large
numbers, i.e. it converges almost surely to  
\be
\avg{A|y}=\frac{\tanh(\bar u+y)-\tanh(\bar u-y)}{2}
\label{Ay}
\ee
(with  $\avg{\ldots|y}$ we denote expected values with respect to the
distribution of $n_i(t)$, at fixed $U^{\pm}$). This in turn implies that \req{dyny} 
is well approximated by
\be
\label{dynydet}
y(t+1)=y(t)(1-\lambda)-\Gamma \avg{A|y(t)}.
\ee
so that, if $y(t)=y$, $A(t)$ has variance
\BE
\avg{\delta A^2|y}&=&\frac{2-\tanh^2(\bar u+y)-\tanh^2(\bar u-y)}{4N}\\
&=&\frac{\cosh^{-2}(\bar u+y)+\cosh^{-2}(\bar u-y)}{4N}
\label{A2y}
\EE
which vanishes as $N\to\infty$.

Equation (\ref{dynydet}) is a nonlinear map with multiplicative noise.
It possesses a trivial fixed point $y^*=0$, which is stable  if 
$|1-\lambda-\Gamma[1-\tanh(\bar u)^2]|<1$
i.e. when
\be
\label{eq:condstab}
\Gamma<(2-\lambda)\cosh^2(\Gamma \eps/\lambda).
\ee
For $\eps=0$, one recovers the stability condition of the original MG,
$\lambda+\Gamma<2$ with $2N$ players \cite{MoCZ}. The role of $\eps$
is then to stabilize the system, as it decreases the right-hand side of
\req{eq:condstab}, independently of its sign. Note that \req{eq:condstab}
can be re-written as 
\be
\eps> \frac{\lambda}{\Gamma}
{\rm acosh}\sqrt{\frac{\Gamma}{2-\lambda}}
\ee
Also note that increasing $\lambda$ drives the system closer to the
instability.

\begin{figure}
 \centerline{\includegraphics*[width=8cm]{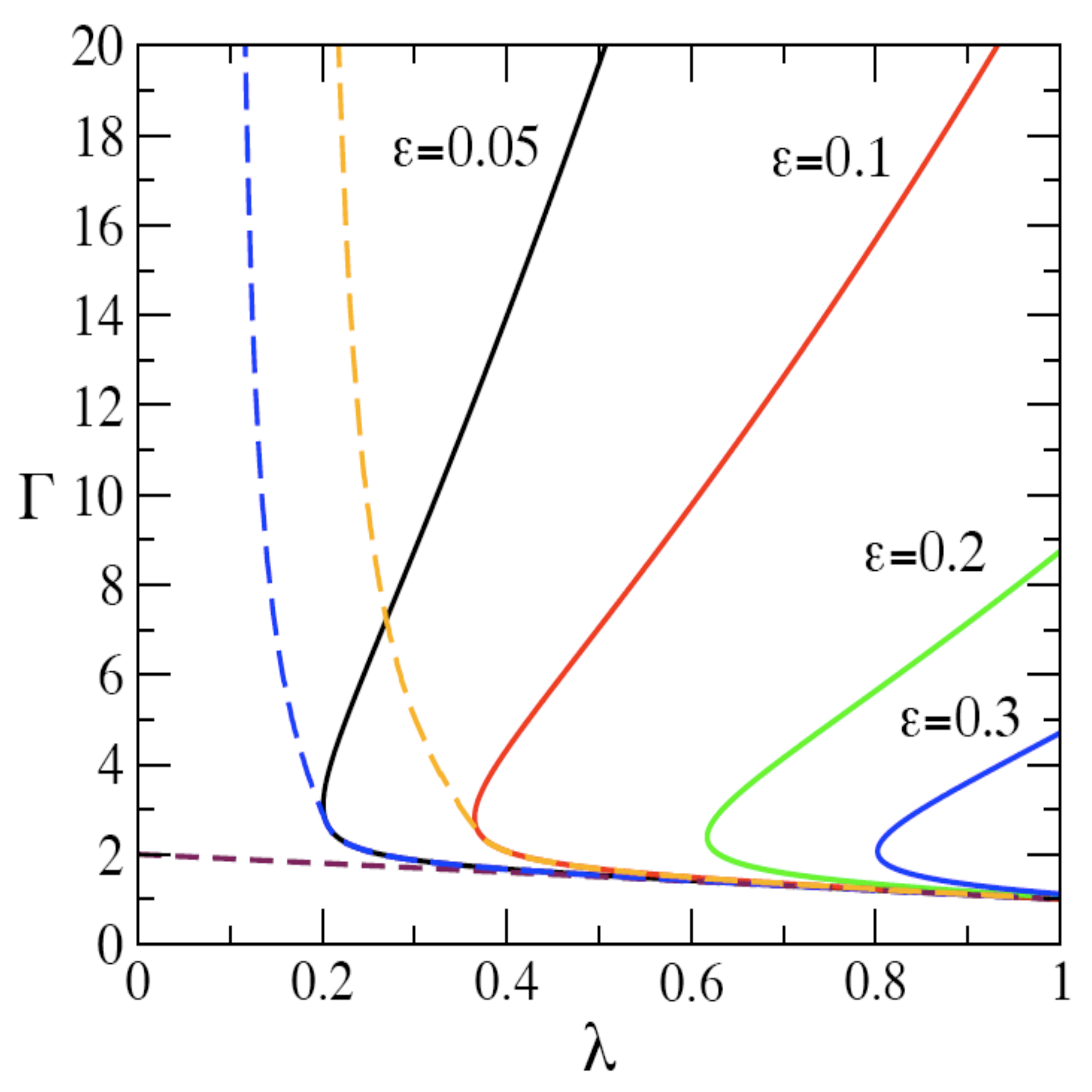}}
 \caption{\label{fig:eps_lambda_Gamma}Stability in the parameter space. The trivial fixed point is stable on the left of the respective $\eps$ lines (continuous curves) and below the $2\Gamma+\lambda=2$ line (dashed line). The stability lines of the oscillating solution (dashed curves) are shown only for $\eps=0.05$ and $\eps=0.1$ for sakes of clarity.}
 \end{figure}

\begin{figure}
 \centerline{\includegraphics*[width=8cm]{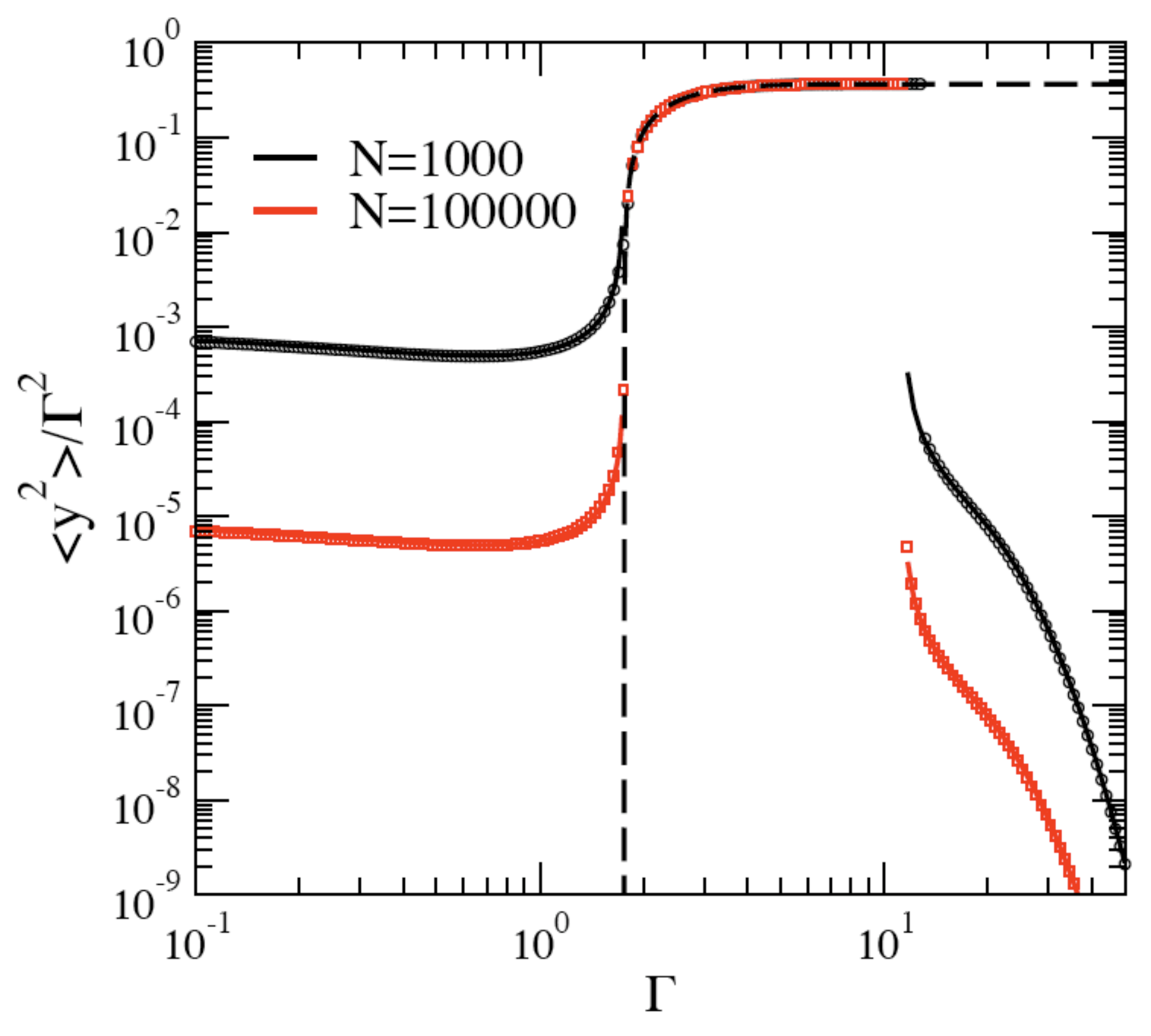}}
 \caption{\label{fighiur}Fluctuations of $y(t)$ for $\epsilon=0.05$, $\lambda=0.35$ as a function of $\Gamma$. In the unstable region (intermediate $\Gamma$), fluctuations are dominated by the
periodic oscillation of $y(t)$, giving rise to a
$\avg{y^2}$ that can be computed analytically (dashed line).}
 \end{figure}

Figure \ref{fig:eps_lambda_Gamma} reports the stability condition of the $y^*=0$ fixed point in the $(\lambda,\Gamma)$ plane (full lines). 
A notable feature of these curves is the existence of two transitions
as $\Gamma$ is varied while $\lambda$ and $\eps$ are kept
constant. 
This is confirmed by the numerical simulations shown in
Fig. \ref{fighiur}, where the fluctuations $\avg{y^2}$ of $y$ around its
fixed point are plotted as a function of $\Gamma$ at fixed $\epsilon$
and $\lambda$, for different values of $N$.

In the region where $y^*=0$ is unstable, a different solution must be considered.
Indeed, besides the trivial fixed point, there is also a period two solution of the form
$y(t)=(-1)^t z^*$, where $z^*$ solves the equation
\be
\label{z}
z=\frac{\Gamma}{2-\lambda}\avg{A|z}=\frac{\Gamma}{2-\lambda}
\frac{\sinh (2z)}{\cosh(2\bar u)+\cosh(2z)}
\ee
This admits again a trivial solution $z^*=0$, which is stable in the same region where the $y^*=0$ solution is. In addition, it can also have a solution $z^*\neq 0$ which is stable in the region where the $y^*=0$ solution in unstable. The stability of this solution may be studied precisely in the same way as before, and it reveals that stability extends beyond the region where the $y^*=0$ solution is unstable, as shown in Fig. \ref{fig:eps_lambda_Gamma}.  In particular, it extends to the region of large $\Gamma$ for $\lambda>2\epsilon/(1+\epsilon)$, where the solution takes the simple asymptotic form $z\simeq 1/(2-\lambda)$ ($\Gamma\gg 1$). This region is limited by a dashed curve in Fig. \ref{fig:eps_lambda_Gamma}, which indicates that the transition form the $z^*\neq 0$ solution to the $y^*=0$ solution is discontinuous in the region where the latter is stable, whereas it is continuous in the region where the fixed point $y^*=0$ is unstable. In the region between the dashed and the solid line both solutions coexist.

So when $\lambda$ is large enough and $\Gamma$ is increased from a small value, we expect a continuous transition, whereas when $\Gamma$ is large enough and $\lambda$ increases from zero across the transition, we find a sharp discontinuous transition and phase coexistence. The latter matches closely the numerical finding of \cite{finitemem} which reported the onset of instabilities occurring for very large times in a more complex version of the present model.

It is worth to comment on the effect of $\epsilon$ on the dynamics.
In the fixed point $y^*=0$, $\epsilon$ controls the average fraction of
agents, thorugh $\bar u$:
\[
\avg{n_i}=\frac{1+\tanh(\bar u)}{2}=\frac{1+\tanh(-\Gamma\epsilon/\lambda)}{2}
\]
This decreases with $\epsilon$ and it vanishes if $\epsilon\gg
\lambda/\Gamma$ is large and positive. On the other hand $\avg{n_i}\to
1$ if $\epsilon$ is large and negative, i.e. all agents play if the
incentive is large enough. This is the main effect of the sign of
$\epsilon$. Indeed both the stability condition as well as the value
of $z$ and the fluctuation properties (see below) are independent
of the sign of $\epsilon$.

\section{Fluctuations in the stochastic system}

In order to explain this rich behavior, it is necessary to study the fluctuations
in the stochastic system.

Let us first notice that $\avg{y^2}\simeq z^2$ 
on the periodic solution $z^*\neq 0$. On this solution, the fluctuations in $A$ are of order one. Indeed they are given by
\[
\avg{A^2}=\avg{A}^2+\avg{\delta A^2}\simeq
\left(\frac{2-\lambda}{\Gamma}\right)^2 z^2+O(N^{-1})
\]
where we have neglected fluctuations $\avg{\delta A^2}\sim
O(N^{-1})$. In brief, the fluctuations are dominated by the periodic oscillation and the  noise has a negligible effect.

Things are a bit more complicated for fluctuations around the $y^*=0$ fixed point. The starting point of our analysis is (\ref{dyny}). We take the square of each side and average over the realizations of the stochastic process. In the stationary state, after rearranging terms, this yields
\be
\label{y2}
[1-(1-\lambda)^2]\avg{y^2}+2\Gamma(1-\lambda)\avg{yA}-\Gamma^2\avg{A^2}=0
\ee
Now, the last two averages take into account both the effect of random choice, for a fixed value of $y$, as described by \req{Ay} and \req{A2y}, and the effect induced by the fluctuations of $y$ itself. 
For the first term, we may write 
\[
\avg{yA}=\int \!dy P(y) y\avg{A|y}\simeq \left.\frac{\partial \avg{A|y}}{\partial y}\right|_{y=0}\avg{y^2}+O(\avg{y^4})
\]
where in the second equality we have expanded the conditional average around the fixed point $y=0$. This was to be expected: because the stochastic noise is of order $1/N$, also fluctuations $\delta y\sim 1/\sqrt{N}$ are likely to be small. Therefore for large $N$ we can safely neglect higher order terms in the above equation. Likewise, we can approximate
\BE
\avg{A^2}&=&\int \!dy P(y) [\avg{\delta A^2|y}+\avg{A|y}^2]\\
&\simeq &
\avg{\delta A^2|0}+
\left[\left.\frac{\partial \avg{A|y}}{\partial y}\right|_{y=0}\right]^2\avg{y^2}+\ldots
\EE
Here, since $\avg{\delta A^2|0}\sim 1/N$, we have neglected
contributions due to fluctuations in $y$, as they only matter to
higher orders in $1/N$. With these,  \req{y2} can easily be solved to yield the fluctuations of $y$:
\be
\label{y22}
\avg{y^2}=\frac{\Gamma^2}{2N\cosh^2(\bar u)
\left[1-\left(1-\lambda-\Gamma/\cosh^2(\bar u)\right)^2\right]}
\ee
Notice that fluctuations diverge as the instability line is
approached, as expected. \req{y22} agrees perfectly with numerical
simulations, as shown in Fig. \ref{fighiur}. 

The approach discussed here can easily be extended, along the lines of
Ref. \cite{Rus}, in order to describe the onset of anomalous
fluctuations beyond the gaussian regime, close to the instability
line.

\section{Conclusion}

We have shown that discounting can have non-trivial consequences on
the dynamics of a system of interacting adaptive agents. From a
na\"ive point of view, this sounds counterintuitive as discounting
decreases the correlation time, hence the strength of non-linear
effects. The interplay with the stochastic nature of the learning
process (due to experimentation) is crucial. Indeed the instability
does not arise (for $\lambda<1$) neither if the noise is too
strong (small $\Gamma$) or if the dynamics is close to being
deterministic ($\Gamma$ large).

We argue that the dynamic instability discussed here is of the same
nature as that which is responsible for the onset of large
fluctuations in the Minority Game discussed in Ref. \cite{finitemem}.
Indeed, the mechanism discussed here is likely of a very general
nature and a similar phase structure may arise in other games with
discounting and experimentation.

\section*{Acknowledgements}

This work was partially supported by EU-NEST grant 516446
ComplexMarkets.

\end{document}